\documentclass[10pt]{article}
\usepackage{geometry}                
\usepackage[parfill]{parskip}
\usepackage{amssymb}

\usepackage{amsmath}

\title{Eigenstates of linear combinations of phase operators}
\author{C. V. Sukumar \\{\em Wadham College,}\\{\em University of Oxford, Oxford OX1 3PN, U.K.}}

\begin{document}
\maketitle
\
\begin{abstract}

The eigenstates of linear combinations of the Susskind and Glogower
phase operators for the harmonic oscillator are constructed. It is
shown that such eigenstates are squeezed states.
\end{abstract}

\centerline{\it PACS INDICES: 03.65, 42.50Dv}
\vfill\eject

It is well known that the coherent states of the harmonic oscillator
~\cite{sc,gl} are eigenstates of the annihilation operator and that the
gaussian squeezed states of the oscillator \cite{yu,hg}  are
eigenstates  of particular linear combinations of the creation and
annihilation operators.  It has recently been shown that the phase
operators of the harmonic oscillator first considered by Susskind and
Glogower \cite{sg,cn} may be used to construct non-Gaussian squeezed
states \cite{s1,s2}  of the oscillator. In this paper it is shown that
the eigenstates of linear combinations of the Susskind and Glogower
phase operators are squeezed states. For simplicity of notation,
Planck's constant $\hbar$ and the frequency of the oscillator are set
equal to 1.

Consider a harmonic oscillator mode described by the creation and
annihilation operators $a^{\dag}$ and $a$ which satisfy
$[a,a^{\dag}]=1$. The Hamiltonian for the oscillator is $H=N+1/2$ in
which $N=a^{\dag}a$ is the number operator. The oscillator energy
eigenstates are denoted by $|n\rangle ,\ n=0,1,2,...$ where $|n\rangle$
is an eigenstate of $N$ with eigenvalue $n$. The coherent state
defined by
\begin{equation}
\Psi_{c} = e^{\eta a^{\dag}-\eta^{\star} a}\ |0\rangle 
\end{equation}
is generated from the vacuum by the displacement operator and is an
eigenstate of $a$ with eigenvalue $\eta$. The squeezed vacuum defined
by
\begin{equation}
\Psi_{s} = e^{(\xi a^{{\dag}^2}-\xi^{\star}a^2)/2}\ |0\rangle 
\end{equation} 
is generated from the vacuum by the squeeze operator. $\Psi_s$ is an eigenstate
 with eigenvalue zero of the operator linear combination defined by
\cite{hg}
\begin{equation}
A= a-\nu a^{\dag} 
\end{equation}
 
where
\begin{equation}
\nu=e^{i\chi} \tanh r\quad ,\quad\ \xi=re^{i\chi}
\end{equation}
It is clear that the magnitude of $\nu$ is restricted to be less than 1.
The eigenstates of A with non-zero eigenvalues are the squeezed and
displaced states obtained from the squeezed vacuum state by the
application of the displacement operator \cite{hg}.

Susskind and Glogower \cite{sg} considered operators of the form
\begin{equation}
P^- =(N+1)^{-1/2} a \quad , \quad\  P^+ =a^{\dag} (N+1)^{-1/2} 
\end{equation}
which have the following properties:
\begin{align}
P^- |n \rangle &=|n-1\rangle , \ n=1,2,\cdots, \notag \\
P^- |0 \rangle &=0, \notag \\
P^+ |n \rangle &=|n+1\rangle ,\ n=0,1,2,\cdots, \notag \\
P^- P^+ |n\rangle &=|n\rangle ,\ n=0,1,2,\cdots, \notag \\
P^+ P^-|n\rangle &=|n\rangle, \ n=1,2,\cdots,\notag \\ 
P^+ P^-|0\rangle &=0 \label{}
\end{align}
The operators $P^{\pm}$ whose classical analogues are $\exp{\mp i\theta}$, where $\theta$ is the phase of the classical oscillator, may then be used to
construct the Hermitian linear combinations
\begin{equation}
C =(P^- + P^+ )/2\quad ,\quad\ S=(P^- - P^+ )/(2i)
\end{equation} 
in terms of which uncertainty relations for phase and number may be
constructed \cite{cn}. The expectation values of C and S for a coherent state
correspond to $\cos \theta$ and $\sin \theta$ respectively when the mean
number of photons in the state is large.

It has recently been shown \cite{s1} that operators of the form
\begin{equation}
R=P^- N \quad ,\quad\ R^{\dag}=N P^+
\end{equation} 
may be used to construct the state
\begin{equation}
\Psi_R =e^{(\beta R^{\dag} - \beta ^{\star} R)} \ |0\rangle 
\end{equation}
$\Psi_R$ may also be written in the form
\begin{equation}
\Psi_R = [1-{|\alpha |}^2] \ \sum_{n=0}^{\infty} \alpha ^n \ |n\rangle
\end{equation}
where
\begin{equation}
\alpha =e^{i\zeta} \tanh \sigma \quad, \quad \beta =\sigma e^{i\zeta}.
\end{equation}
It has been shown that $\Psi_R$ is a squeezed state. Using equations
(6) and (10) it can be seen that $\Psi_R$ is an eigenstate of
$P^-$ with eigenvalue $\alpha$. 

Similarly operators of the form
\begin{equation}
S=P^{-^2} N \quad ,\quad\ S^{\dag}=N P^{+^2}
\end{equation} 
may be used to construct the state
\begin{equation}
\Psi_S =e^{(\gamma S^{\dag} - \gamma ^{\star} S)}\  |0\rangle
\label{2}
\end{equation}
which may also be written in the form
\begin{equation}
\Psi_S = [1- {|\delta |}^2] \ \sum_{n=0}^{\infty} \delta ^{n}
\ |2n\rangle
\end{equation} 
where 
\begin{equation}
\delta =e^{i\phi} \tanh \rho \ \quad ,\ \ \gamma =(\rho /2) e^{i\phi} 
\end{equation}
$\Psi_S$ has also been shown \cite{s2} to be a squeezed state. Using equations (6)
and (14) it is readily established that $\Psi_S$ is an eigenstate of $P^{-^2}$
with eigenvalue $\delta$. It is simple to verify that $\Psi_S$ is also
an eigenstate of the operator
\begin{equation}
B=(P^- -\delta P^+) 
\end{equation}
with eigenvalue zero. The squeezed states $\Psi_R$, $\Psi_S$ and the
operator $B$ are in a sense the phase analogues of the states 
$\Psi_c$, $\Psi_s$ and the operator $A$. Such an analogy leads us to consider the question
of eigenstates of $B$ with non-zero eigenvalue which would  be
analogous to the squeezed and displaced states which are eigenstates of
$A$ with non-zero eigenvalue.

Let
\begin{equation}
\Psi=\sum_{n=0}^{\infty} C_n \ |n\rangle
\end{equation}
be a solution of 
\begin{equation}
[P^- - \delta P^+] \ \Psi = \mu \ \Psi \ .
\end{equation}
Then $C_n$ must satisfy
\begin{equation}
C_1 = \mu C_0 \ ,\ \  C_{n+1} - \delta C_{n-1} = \mu C_n,\ n=1,2,\cdots  \\
\end{equation}
For the special case when $\delta = \mu ^2$
a solution to the above recursion relations is given by $C_n = \mu ^n F_n$ where $F_n$ is the $n^{th}$ Fibonacci number. A procedure analogous to
that used for finding closed analytic expression for the Fibonacci
sequence may be used to solve the recursion relations given by
equation (19) in the general case when $\delta$ and $\mu$ are free
parameters. The solution for $C_n$ may be given in the form
\begin{equation}
C_n =C_0\ \delta ^{n/2}\ [ p_1^{n+1} - p_2^{n+1} ]/[ p_1 - p_2 ]
\end{equation}
where
\begin{align}
p_1 &= [\lambda + \sqrt{\lambda ^2 +4}]/2,\notag \\ p_2 &=[\lambda -
\sqrt{\lambda ^2 +4}]/2 \notag \\
\lambda &=\mu/\sqrt{\delta} 
\end{align}
In deriving the above solution it has been assumed that $\lambda\not= 2i$.
In the limiting case that $\lambda\rightarrow 2i$, it is clear that $p_1\rightarrow p_2$ and L'Hospital's rule may be used to
simplify equation (20) to the form 
\begin{equation}
C_n\ =\ C_0 \ (n+1)\ i^n\ \delta ^{n/2}\ , \ \mu =2i\sqrt{\delta}
\end{equation}
In rest of this paper we will consider the solution given by
equations (20) and (21) and restrict the analysis to real values of $\delta$ and
$\mu$.

The free time evolution of a state which at time t=0 has the form given  by equation (17) is
\begin{equation}
\Psi (t) \ =\ \sum_{n=0}^{\infty} \ e^{-i(n+1/2)t} \ C_n \
|n\rangle .
\end{equation}
The variances of the Hermitian linear combinations
\begin{equation}
q=(a+a^{\dag})/\sqrt{2} \ \ ,\ \ p=-i(a-a^{\dag})/\sqrt{2} 
\end{equation}
for the state $\Psi (t)$ when $C_n$ are chosen to be real can be shown to be given by
\begin{align}
Var(q) &=1/2\ -(F_1-F_2)+2(F_1-F_3^2)\cos ^{2} t   \\
Var(p) &=1/2\ -(F_1-F_2)+2(F_1-F_3^2)\sin ^{2} t   
\end{align}
where
\begin{align}
F_1 &=\sum_{n=0}^{\infty}C_nC_{n+2}\sqrt{(n+1)(n+2)} \\
F_2 &=\sum_{n=0}^{\infty} C_n^2 \ n   \\ 
F_3 &=\sum_{n=0}^{\infty} C_nC_{n+1} \sqrt{n+1}
\end{align}
A state $\Psi$ is said to be squeezed when Var(p) or Var(q) can dip
below the value 1/2. A sufficient condition for squeezing is that
$F_1\ >\ F_2$.

It is easy to show that when $C_n$ is given by equations (20) and (21) then
\begin{equation}
C_n C_{n+2} \ -\ C_{n+1}^{2} \ =\ C_0^2\ \delta ^{n+1} \ (-)^n\ \ ,\ \ C_n C_{n+2}\ >\ 0 \ 
\end{equation}
The above conditions together with the property that 
$\sqrt{(n+1)(n+2)}>(n+1) $ for any $n$ and the expression
\begin{equation}
\sum_{n=0}^{\infty} (-)^{n} \delta^{n+1} \ (n+1)\ =\ (1+\delta )^{-2}
\end{equation}
may be used to establish that for the state defined by equations (17),(20)
and
(21), $(F_1\ -\ F_2)\ >\ 0$ and therefore it is a squeezed state.
We have shown that the eigenstates of $B$ defined by equation (16) are
given by equations
(17),(18),(20) and (21) and that these states are squeezed states. It is
interesting to note that in particular a linear superposition of
number states of the form given by equation (17) with the coefficients
$C_n$ given by a Fibonacci sequence is a squeezed state.

The state $\Psi_R$ defined by equation (10) has the form of a thermal
superposition. Hence we may interpret the squeezed state $\Psi$ defined
by equations (17) and (20) as a linear of superposition of two thermal
states each of which is a squeezed state. Multiplication of both sides
of equation (18) from the left by $P^-$ and use of the property that
$P^-P^+=1$ leads to 
\begin{equation}
[P^{-^2}\ -\ \mu P^-]\ \Psi\ =\ \delta \ \Psi 
\end{equation}
which shows that $\Psi$ is also an eigenstate of the operator
$[P^{-^2}-\mu P^-]$ with eigenvalue $\delta$.

The structure of equation (30) suggests that a linear superposition of
the form
\begin{equation}
\Phi =\sum_{n=0}^{\infty}\ C_n/\sqrt{n!}\ \ |n\rangle 
\end{equation}
with $C_n$ given by equations (20) and (21)  will lead to
\begin{equation}
F\equiv (F_1-F_2)\ =\ C_0^2 \ \delta \ e^{-\delta} 
\end{equation}
The positivity of $F$ for positive $\delta$ guarantees that $\Phi$ is
also a squeezed state. It can be established that $\Phi$ is a solution
of the eigenvalue equation 
\begin{equation}
(a^2\ -\ \mu a)\ \Phi \ =\ \delta \ \Phi \ .
\end{equation}
Hence the squeezed state defined by equations (20),(21) and (33) is an
eigenstate of $(a^2-\mu a)$ with positive eigenvalue $\delta$. When
written in the form
\begin{equation}
\Phi = C_0p_1/(p_1-p_2)\sum_{n=0}^{\infty}{(p_1\sqrt{\delta})}^n /\sqrt{n!}\
\ |n\rangle\
-\ C_0p_2/(p_1-p_2)\sum_{n=0}^{\infty}{(p_2\sqrt{\delta})}^n /\sqrt{n!}\
\ |n\rangle
\end{equation}
it is clear that $\Phi$ is a linear superposition of of two coherent
states with displacement parameters $(p_1\sqrt\delta)$ and
$(p_2\sqrt\delta)$ respectively. The construction of squeezed states by
the superposition of coherent states and thermal states has been the
subject of some recent discussion \cite{bk}. It is clear that equations (33)
and (36) for the coherent state superposition $\Phi$ are analogous to
equations (17) and (20) for the thermal state superposition $\Psi$.

In this paper we have shown that the eigenstates of linear combinations
of the Susskind-Glogower phase operators may be readily constructed and
that such states are squeezed states. We have shown that there is a
great deal of similarity between the properties of the eigenstates of
operator linear combinations generated from the operator sets
$(P^-,P^+)$ and $(a,a^{\dag})$.

\end{document}